\documentclass[11pt,a4paper]{article}
\pdfoutput=1
\usepackage{jheppub}

\usepackage{url}

\allowdisplaybreaks[2]

\def\cB{\mathcal{B}}
\def\cC{\mathcal{C}}

\def\cO{\mathcal{O}}

\def\mint{\int_{-\infty}^\infty\!\cdots\!\int_{-\infty}^\infty}

\newcommand{\be}{\begin{equation}}
\newcommand{\ee}{\end{equation}}
\newcommand{\ba}{\begin{aligned}}
\newcommand{\ea}{\end{aligned}}

\def\({\left(}
\def\){\right)}

\DeclareMathOperator{\im}{Im}

\preprint{RUP-19-18}

\title{Quasinormal modes of black holes and Borel summation}

\author{Yasuyuki Hatsuda}

\affiliation{Department of Physics, Rikkyo University, Toshima, Tokyo 171-8501, Japan}

\emailAdd{yhatsuda@rikkyo.ac.jp}

\abstract{
We propose a simple and efficient way to compute quasinormal frequencies of spherically symmetric black holes.
We revisit an old idea that relates them to bound state energies of anharmonic oscillators by an analytic continuation.
This connection enables us to achieve remarkable high-order computations of WKB series by Rayleigh--Schr\"odinger perturbation theory. The known WKB results are easily reproduced.
Our analysis shows that the perturbative WKB series of the quasinormal frequencies turn out to be Borel summable divergent series both for the Schwarzschild and for the Reissner--Nordstr\"om black holes. Their Borel sums reproduce the correct numerical values.
}

\begin{document}

\maketitle

\renewcommand{\thefootnote}{\arabic{footnote}}
\setcounter{footnote}{0}
\setcounter{section}{0}

\section{Introduction}
In black hole perturbation theory, characteristic oscillatory modes appear.
Because of emission of gravitational radiation, these oscillations decay, and thus are called quasinormal modes.
The quasinormal modes play a central role at the final stage, ringdown phase, of coalescence of two black holes,
and have a direct connection with the recent observation of gravitational waves~\cite{abbott2016, abbott2016a, abbott2016b}.
It is an important task to compute the quasinormal frequencies for various black holes as precise as possible.
Unfortunately, it is almost impossible for us to follow a huge number of references on this subject. We refer to a few comprehensive reviews~\cite{kokkotas1999, berti2009, konoplya2011}.

The purpose of the present work is to develop a widely applicable method to compute the quasinormal frequencies of spherically symmetric black holes. 
We combine a few new ideas with some known ones in various fields.
The method that we propose here is simple and efficient.
Anyone who has basic skills on Mathematica can compute the accurate quasinormal frequencies for a wide class of black holes from now on!
We explicitly demonstrate it for two simple examples: the Schwarzschild black hole and the Reissner--Nordstr\"om black hole.%
\footnote{Mathematica codes for these examples are attached to the source file on arXiv.} 
It is very straightforward to generalize our method to many other cases.

It is well-known that the WKB approach, initiated by Mashhoon~\cite{mashhoon1983} and by Schutz and Iyer~\cite{schutz1985},%
\footnote{We should note that in the WKB approach, one does not necessarily have to expand a potential around its maximum. The general treatment leads to Bohr--Sommerfeld-like quantization conditions~\cite{froman1992}, which are more accurate than the approach in~\cite{schutz1985} for high overtone numbers in general.} 
can be applied to many cases. In~\cite{iyer1987}, Iyer and Will computed the third order correction, and Konoplya extended it to the sixth order~\cite{konoplya2003}. Recently, the computation up to the 13th order has been done by Matyjasek and Opala~\cite{matyjasek2017} (see also~\cite{konoplya2019}). While these great computations actually improved approximate values of the quasinormal frequencies of black holes, it seems hard to answer more fundamental questions: Are the WKB series convergent or divergent? Do they receive nonperturbative corrections? Can we reconstruct the exact quasinormal frequencies from their WKB series? Of course, these questions are all interrelated.
To answer them, much higher-order data are desirable.
The high-order data also provide us more accurate numerical values as a result.
In this work, we propose a simple way to do so.
It easily reproduces the previous WKB results.
We emphasize that the application range of our approach is as wide as that of the WKB approach.

We revisit an old idea originally proposed by Blome, Ferrari and Mashhoon~\cite{blome1984, ferrari1984, Ferrari:1984zz}.
It is explained in those papers that the quasinormal frequencies of black holes are related to the bound state energies of anharmonic oscillators by an analytic continuation.
We find that this connection allows us to use a powerful technique, \textit{\`a la} Bender and Wu~\cite{bender1969}, in quantum mechanics.
Surprisingly, no one has ever applied this famous technique to the computation of the quasinormal frequencies, though it has been known for half a century!
Recently, from another motivation in high-energy physics, the Bender--Wu method has been beautifully packaged in Mathematica by Sulejmanpasic and \"Unsal~\cite{sulejmanpasic2018}.
To the author's knowledge, this is now the best tool to look into perturbative series deeply in quantum mechanics. 
Based on these excellent works, we show that the Bender--Wu method is indeed greatly useful in the computation of the quasinormal modes.
By using this method, we can compute the perturbative WKB expansion to extremely high orders.
We have reached the 200th order for the Schwarzschild and for the Reissner--Nordstr\"om black holes, and one can easily go beyond it, depending on one's CPU power and time constraint.%
\footnote{The 50th order evaluation for a given overtone number will be done in 15-30 seconds, and the 200th order in 10-15 minutes on recent home computers.} 
Note that our approach also gives the spectrum of bound states in an associated eigenvalue problem as a bonus. 

Such high-order data are useful to clarify the convergence or the divergence of the WKB series at the very quantitative level.
We find that the WKB series is almost surely divergent, 
\textit{i.e.}, its radius of convergence is just zero.
The same result has been observed in~\cite{matyjasek2017, konoplya2019}. Our result gives much stronger evidence.
This result is not surprising because in quantum mechanics, in quantum field theories and in string theory, perturbative series are usually divergent. The Borel summation method tells us a lot of important information on exact functions from their perturbative series.
In the examples in this paper, the WKB series of the quasinormal frequencies are always Borel summable, and it strongly implies that  they do not receive any nonperturbative corrections.
As a consequence, we conclude that the Borel summation of the WKB series gives the exact quasinormal frequency.

The organization of this paper is as follows.
In the next section, we simply re-derive the WKB results in our approach. We map the problem to Rayleigh--Schr\"odinger perturbation theory in quantum mechanics. 
Section~\ref{sec:Borel} is the main part of this paper. 
We show the Borel analysis for the Schwarzschild and for the Reissner--Nordshtr\"om black holes.
The singularity structure of the Borel transform reveals that the perturbative WKB series are Bore summable divergent series. It implies that the WKB series do not receive nonperturbative corrections.
We can easily perform the Borel summation, and it agrees with the known numerical data with remarkable accuracy.
We comment on some future directions in section~\ref{sec:summary}.
For the reader who is not familiar with the Borel summation method, we give a brief review in appendix~\ref{app:Borel}.

\section{Semiclassical perturbative expansions}\label{sec:semiclassical}
In~\cite{blome1984, ferrari1984, Ferrari:1984zz}, quasinormal frequencies of black holes are related to bound state energies.
There, the relationship was applied to exactly solvable potentials.
We stress that this idea is quite general, and it is not restricted to the special application.
Here we revisit this relation in a refined way.
The similar computation is also found in~\cite{zaslavskii1991}, but our approach is much more efficient.
The obtained result is directly compared with the WKB results in~\cite{schutz1985, iyer1987, konoplya2003}.

\subsection{Basic idea}
\begin{figure}[tbp]
\begin{center}
\includegraphics[width=0.8\linewidth]{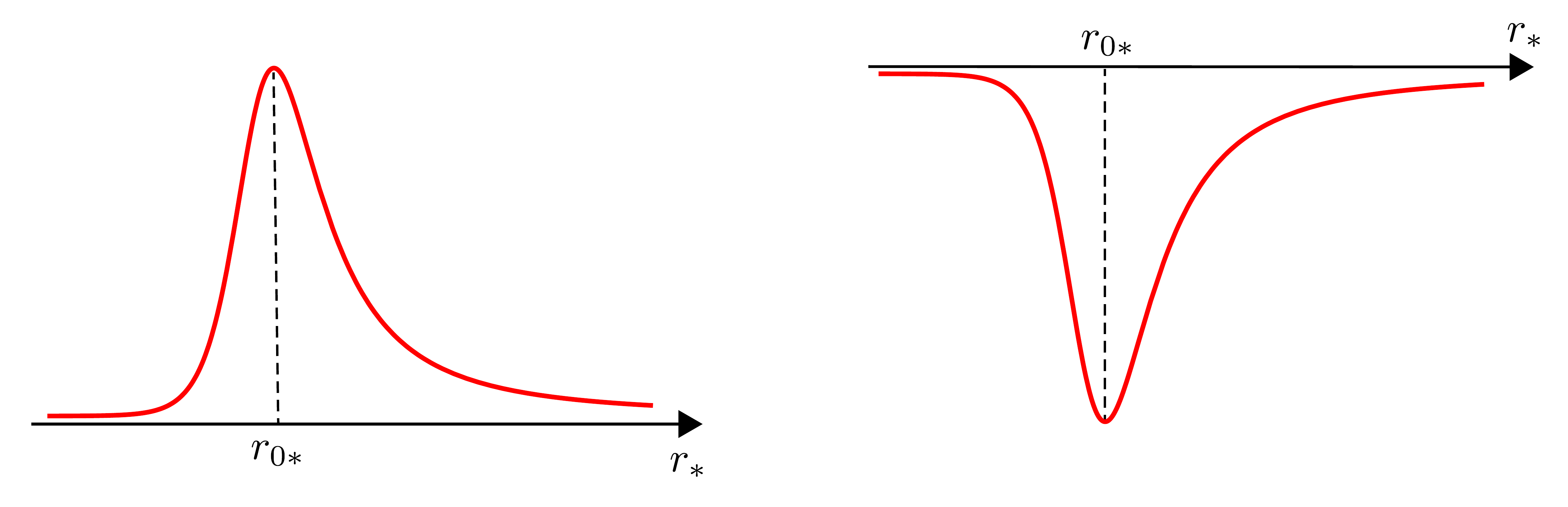}
\caption{(Left) A typical shape of the potential $V(r_*)$. (Right) The inverted potential $-V(r_*)$ usually has bound states.}
\label{fig:potential}
\end{center}
\end{figure}

Throughout this paper, we focus on spherically symmetric black holes. Black hole perturbation theory leads to the following radial master equation: 
\begin{equation}
\begin{aligned}
\(\epsilon^2 \frac{d^2}{dr_*^2}+\omega^2-V(r_*) \)\phi(r_*)=0,
\end{aligned}
\label{eq:Sch-eq}
\end{equation}
where $r_*$ is the tortoise variable. 
The black hole horizon is at $r_*=-\infty$, and the spacial infinity is at $r_*=\infty$.
We have introduced a formal parameter $\epsilon$ that characterizes the WKB series. 
In the language of quantum mechanics, it of course corresponds to the Planck constant, and thus we often refer to
the expansion around $\epsilon=0$ as the semiclassical expansion.
Usually, $\epsilon$ is set to be unity.
Typically, the potential $V(r_*)$ has a shape shown in the left of figure~\ref{fig:potential},
and it has a global maximum. Our procedure, however, is not restricted on such typical potentials.

In parallel, let us consider the Schr\"odinger equation with the \textit{inverted} potential:
\begin{equation}
\begin{aligned}
\( -\hbar^2 \frac{d^2}{dr_*^2}-V(r_*) \)\psi(r_*)=E \psi(r_*).
\end{aligned}
\label{eq:Sch-eq-inv}
\end{equation}
It is clear that the inverted potential $-V(r_*)$ has the minimum, shown in the right of figure~\ref{fig:potential}, and usually has bound states for $\hbar>0$ and $E<0$.
We denote its bound state energy by $E_n^\text{BS}(\hbar)$ ($n=0,1,2,\dots$).

Now we consider the analytic continuation of $\hbar$. If setting $\hbar=i\epsilon$, the Schr\"odinger equation \eqref{eq:Sch-eq-inv} with $E=-\omega^2$ formally coincides with \eqref{eq:Sch-eq}.
Therefore, it is expected that the quasinormal frequency $\omega_n^\text{QNM}$ at $\epsilon=1$ is simply related to the bound state energy $E_n^\text{BS}$ at $\hbar=i$ by
\begin{equation}
\begin{aligned}
(\omega_n^\text{QNM})^2=-E_n^\text{BS}(\hbar=i).
\end{aligned}
\label{eq:QNM-BS}
\end{equation}
This is the key equation in our analysis.
Of course, to prove (or disprove) this optimistic guess, we have to carefully see how the boundary conditions on both sides are related by the analytic continuation of $\hbar$. 
Roughly speaking, in the bound state problem, the exponentially decaying solution in $r_* \to \infty$ behaves as $e^{-\sqrt{-E}\, r_*/\hbar}$, and it is analytically continued to the outgoing solution $e^{+i\omega r_*}$, which is the boundary condition imposed at the spacial infinity for the quasinormal modes. Similarly, the decaying solution $e^{+\sqrt{-E}\, r_*/\hbar}$ in $r_* \to -\infty$ is continued to the ingoing mode $e^{-i\omega r_*}$ at the horizon.
Therefore both the boundary conditions seem to be related appropriately by the analytic continuation.

However, this argument is quite intuitive. There is a possibility that the decaying solution leads to a linear combination
of the outgoing and the ingoing solutions after the analytic continuation due to the Stokes phenomenon.
This phenomenon is invisible as far as one considers asymptotic solutions.
The Stokes phenomenon in the WKB method was formulated in the seminal work~\cite{voros1983} of Voros, and now known as the exact WKB analysis. 
We expect that the relation \eqref{eq:QNM-BS} is rigorously (dis)proved by the exact WKB analysis,
but it is beyond the scope of this work. 
Note that the Stokes phenomenon is also important to derive the high overtone asymptotic behavior of the quasi-normal frequencies \cite{motl2003, andersson2004, cardoso2004a, natario2004}.
We currently assume the relation \eqref{eq:QNM-BS} without any rigorous proofs.  We will check it by comparing the results obtained in this way with the known ones. This is the main goal of this paper.

Let us comment on the difference from the original proposal in~\cite{blome1984, ferrari1984, Ferrari:1984zz}.
The authors in~\cite{blome1984, ferrari1984, Ferrari:1984zz} considered the analytic continuation of the radial coordinates $r$ and $r_*$. To get the inverted potential,
one has to do the further analytic continuation of other parameters (mass, charge, etc.) in the potential simultaneously, while the Planck constant is fixed.
Here, we rather analytically continue only the Planck constant. Though these two complementary procedures seem equivalent, ours looks much simpler.

\subsection{Re-deriving WKB results}

We start with the Taylor expansion of the inverted potential:
\begin{equation}
\begin{aligned}
-V(r_*)=V_0+\sum_{k=2}^\infty V_k\;  (r_*-r_{0*})^k ,
\end{aligned}
\end{equation}
where the inverted potential has the minimum at $r_*=r_{0*}$.
For our purpose, it is more useful to define $r_*-r_{0*}=\sqrt{\hbar}\, x$,
and rewrite the Schr\"odinger equation \eqref{eq:Sch-eq-inv} as
\begin{equation}
\begin{aligned}
\(-\frac{1}{2}\frac{d^2}{dx^2}+\frac{V_2}{2}x^2+V_\text{int}(x) \)\psi(x)=\epsilon \psi(x),
\end{aligned}
\label{eq:Sch-eq-2}
\end{equation}
where
\begin{equation}
\begin{aligned}
\epsilon:=\frac{E-V_0}{2\hbar},\qquad V_\text{int}(x):=\frac{1}{2}\sum_{k=3}^\infty \hbar^{k/2-1}V_k x^k.
\end{aligned}
\label{eq:V-int}
\end{equation}
We regard the equation \eqref{eq:Sch-eq-2} as an anharmonic oscillator with an infinite number of interaction terms.
In this viewpoint, the Planck constant is set to be unity, and $\hbar$ plays the role of the coupling constant of the interactions.

The standard (but boring) textbook-like method in perturbation theory leads to the expansion around $\hbar=0$ order by order in principle~\cite{zaslavskii1991}.
Instead, a much  more economical way is known~\cite{bender1969}.
We do not repeat the argument in ~\cite{bender1969}, but mention that
there is a useful Mathematica package%
\footnote{The correctness of this package has been confirmed for various quantum mechanical models studied in a recent revival of resurgence theory \cite{dunne2016, kozcaz2018, codesido2017a, codesido2019}.}~\cite{sulejmanpasic2018} to compute the perturbative expansion of the spectrum for a given potential by this method. See these papers in detail.
With the help of this Mathematica package, we easily get the following perturbative expansion for the eigenvalue equation \eqref{eq:Sch-eq-2}:
\begin{align}
\epsilon_n^\text{pert}&=\sqrt{V_2}\alpha+\hbar\left[ -\frac{V_3^2}{64V_2^2}(7+60\alpha^2)+\frac{3V_4}{16V_2}(1+4\alpha^2) \right] \notag \\
&\quad+\hbar^2 \alpha \biggl[ -\frac{15V_3^4}{1024V_2^{9/2}}(77+188\alpha^2)+\frac{9V_3^2 V_4}{128V_2^{7/2}}(51+100\alpha^2)-\frac{V_4^2}{64V_2^{5/2}}(67+68\alpha^2) \notag \\
&\qquad\qquad-\frac{5V_3V_5}{32V_2^{5/2}}(19+28\alpha^2)+\frac{5V_6}{16V_2^{3/2}}(5+4\alpha^2) \biggr]
+\cO(\hbar^3),
\label{eq:WKB-1}
\end{align}
where $\alpha=n+1/2$.
Clearly, the first term is due to the harmonic potential with the frequency $\sqrt{V_2}$.
The second term comes from the cubic and the quartic interactions.
As in \eqref{eq:V-int}, the interaction term $x^k$ leads to the contribution with order $\hbar^{k/2-1}$.
If one wants to know the perturbative series of $\epsilon_n$ up to the $\ell$-th order, one needs to compute the Taylor expansion up to the $2(\ell+1)$-th order.

This result should be compared to the WKB result in~\cite{iyer1987}.
To do so, we use the relation $Q(r_*)=-E-V(r_*)$ with $\omega^2=-E$, and compare the Taylor expansions on both sides.
We find
\begin{equation}
\begin{aligned}
Q_0=-E+V_0=-2\hbar \epsilon,\qquad
\frac{Q_0^{(k)}}{k!}=V_k.
\end{aligned}
\label{eq:Taylor-comparison}
\end{equation}
Using these relations and setting $\hbar=i$, one finds that our result \eqref{eq:WKB-1} is in agreement with (1.5a) and (1.5b) in~\cite{iyer1987}. 

Even if the function $Q(r_*)$ cannot be divided into the simple form $Q(r_*)=\omega^2-V(r_*)$, we can still regard the constant term $Q_0$ as the ``energy'' in the Schr\"odinger equation. In this case, the extremal point $r_{*0}$, as well as the Tayler coefficients $Q_0^{(k)}$, depends on $\omega$.

\section{Borel analysis}\label{sec:Borel}
The greatest advantage of our approach is that one can push the semiclassical perturbative computation to very high orders.
This high-order computation helps us to understand the analytic structure of $E_n^\text{BS}(\hbar)$ and also $\omega_n^\text{QNM}(\epsilon)$.
Here we show explicit computations for a few examples.
We note that our method is widely applicable for many other cases.

\subsection{Schwarzschild black hole}
For the Schwarzschild black hole, the Regge--Wheeler potential is universally given by
\begin{equation}
\begin{aligned}
V(r_*)=\(1-\frac{1}{r}\)\( \frac{l(l+1)}{r^2}+\frac{1-s^2}{r^3} \) \quad (s=0,1,2),
\end{aligned}
\end{equation}
where the tortoise variable is given by $r_*=r+\log(r-1)$. We have set the Schwarzschild mass to be $2M=1$.
The cases of $s=0,1,2$ correspond to scalar, electromagnetic and gravitational perturbations, respectively.

We here show the explicit computation for the gravitational perturbation ($s=2$) with $l=2$.
The other cases are completely the same.
In this case, the (inverted) potential takes the minimal value at
\begin{equation}
\begin{aligned}
r_{0*}=r_0+\log(r_0-1),\qquad r_0=\frac{27+3\sqrt{17}}{24}.
\end{aligned}
\end{equation}
It is straightforward to compute the Taylor expansion of the potential around $r_*=r_{0*}$.
To make it easy to see the result, we show approximate numerical coefficients:
\begin{equation}
\begin{aligned}
-V(r_*)=-0.605147+0.0793553 \delta_*^2-0.0134245 \delta_*^3-0.00638133 \delta_*^4+0.00263418 \delta_*^5\\
+0.000160259 \delta_*^6-0.000300182 \delta_*^7+0.0000425323 \delta_*^8+\cO(\delta_*^9),
\end{aligned}
\end{equation}
where $\delta_*:=r_*-r_{0*}$. Of course, one can do this computation by keeping the coefficients analytic.
From a practical point of view, it is enough to compute these coefficients numerically, kept sufficiently accurate.
Recall that in order to get the $\ell$-th order perturbative correction to $\epsilon_n$, the Taylor expansion of the potential up to the $2(\ell+1)$-th order
is needed.
In the current case, we can compute it up to any desired order (at least numerically).
Once the Taylor expansion of the potential is known, the Mathematica package in~\cite{sulejmanpasic2018} automatically computes the perturbative series of $\epsilon_n$.
For example, the perturbative series of the ground state and the first excited state energies to the fifth order are given by
\begin{equation}
\begin{aligned}
\epsilon_0^\text{pert}=0.14085-0.0399931 \hbar&+0.00768009\hbar^2-0.000617727\hbar ^3 \\
&-0.000324705\hbar^4+0.000181562\hbar ^5+\cO(\hbar^6), \\
\epsilon_1^\text{pert}=0.422551-0.214274 \hbar&+0.0420492 \hbar ^2-0.000599094 \hbar ^3\\
&-0.0016814 \hbar ^4+0.00136219 \hbar ^5+\cO(\hbar^6), %\\
%\epsilon_2^\text{pert}=0.704252 - 0.562837 \hbar&+0.133445 \hbar ^2-0.0209528 \hbar ^3\\
%&-0.000561438 \hbar ^4+0.0033168 \hbar ^5+\cO(\hbar^6),
\end{aligned}
\end{equation}
To get the quasinormal modes, we finally use the relation \eqref{eq:QNM-BS}, \textit{i.e.}, $\omega_n^2=-(V_0+2\hbar \epsilon_n)$, and set $\hbar=i$. The perturbative series of $\epsilon_n$ up to $\hbar^5$ lead to the following numerical values:
\begin{equation}
\begin{aligned}
\omega_0(s=2,l=2) &\approx 0.747239-0.177782 i , \\
\omega_1(s=2,l=2) &\approx 0.692593-0.546960 i , %\\
%\omega_2(s=2,l=2) &\approx 0.597040 - 0.955121i , 
\end{aligned}
\end{equation}
where we have chosen a branch such that $\im \omega_n<0$.
These values are compared with the sixth order WKB approximation in~\cite{konoplya2003}, and one finds
a precise agreement. Note that from $\omega_n^2=-(V_0+2\hbar \epsilon_n)$, the fifth order correction to $\epsilon_n$ corresponds to the sixth order to $\omega_n^2$. 

Let us write the all-order perturbative expansion of the energy as
\begin{equation}
\begin{aligned}
\epsilon_n^\text{pert}=\sum_{k=0}^\infty \epsilon_n^{(k)}  
\hbar^k.
\end{aligned}
\label{eq:E-pert}
\end{equation}
How can we get the exact quasinormal frequency from this perturbative series?
The answer is not so simple. It turns out that we need a technique in asymptotic analysis.

\begin{figure}[tbp]
\begin{center}
\includegraphics[width=0.4\linewidth]{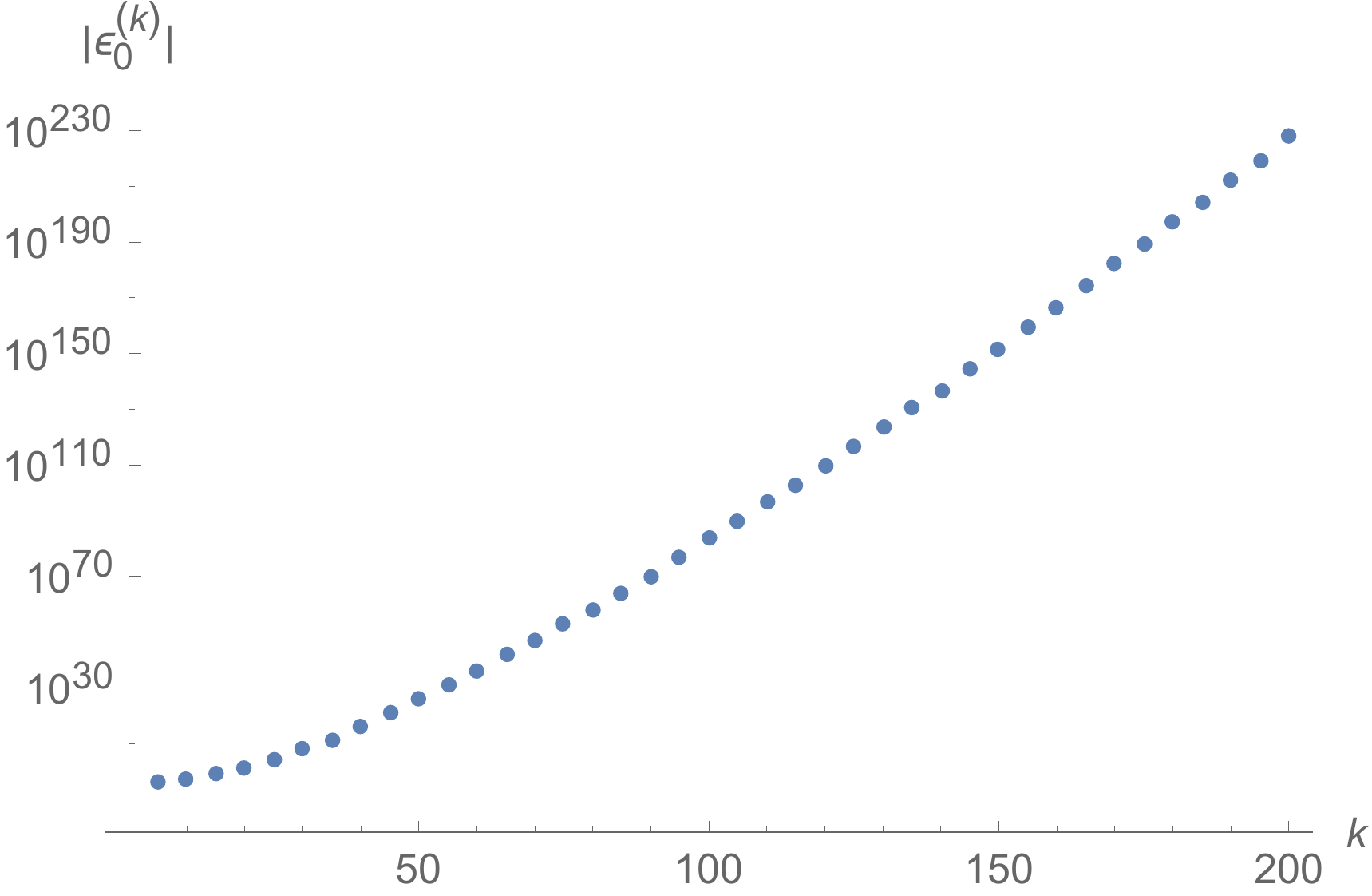}
\caption{The large order behavior of $|\epsilon_0^{(k)}|$ for $(s,l)=(2,2)$.}
\label{fig:coeff}
\end{center}
\end{figure}

We have computed $\epsilon_n^{(k)}$ ($n=0,1,2$) up to $k=200$. Everything can be put on Mathematica. 
The large order behavior of the absolute values of the ground state coefficients is shown in figure~\ref{fig:coeff}.
The same behavior is also observed for excited energies.
This terribly growing behavior strongly suggests that \eqref{eq:E-pert} is a divergent series, as in most quantum mechanical models.
This fact is clearly confirmed by looking at singularities of the Borel transform, as we will show below.

Note that the divergent series \eqref{eq:E-pert} works approximately up to a certain finite order $k_0$.
Beyond this order, the approximation by \eqref{eq:E-pert} gets worse.
If one regards \eqref{eq:E-pert} as an approximate expression, one should truncate all the terms beyond the best order $k_0$. 
This is well-known as the optimal truncation. This behavior is crucially different from convergent series. In convergent series, the approximation (inside the convergence circle) gets better and better by computing higher and higher order corrections. This is not true for divergent series.
In this sense, it is very important to know whether a considered perturbative series is convergent or divergent.

The perturbative series beyond the optimal order does not work as an approximate expression any more.
The reader might think that the higher-order corrections in the perturbative series are useless.
This is not the case. There are some ways to improve the approximation.
One way is to use Pad\'e approximants, as in~\cite{matyjasek2017}.
However, it seems difficult in this way to understand the analyticity property of the WKB series, such as the Stokes phenomenon.
Moreover, to the author's knowledge, the convergency is unclear when Pad\'e approximants are applied to divergent series.
For these reasons, we here use another important method, well-known as \textit{Borel summation}. 
The Borel summation is a basic tool in analysis of divergent series.
We review this method in appendix~\ref{app:Borel}.

As in appendix~\ref{app:Borel}, we define the Borel transform of \eqref{eq:E-pert} by
\begin{equation}
\begin{aligned}
\cB[\epsilon_n^\text{pert}](\zeta):=\sum_{k=0}^\infty \frac{\epsilon_n^{(k)}}{k!} \zeta^k.
\end{aligned}
\end{equation}
We denote its analytic continuation by $\cB^{\cC}[\epsilon_n^\text{pert}](\zeta)$.
Then the Borel sum is given by the Laplace transform:
\begin{equation}
\begin{aligned}
\epsilon_n^\text{pert,Borel}(\hbar)=\int_0^\infty d\zeta\, e^{-\zeta} \cB^{\cC}[\epsilon_n^\text{pert}](\hbar \zeta),
\end{aligned}
\label{eq:E-Borel-sum}
\end{equation}
The Borel summed quasinormal frequency is finally defined by
\begin{equation}
\begin{aligned}
(\omega_n^\text{Borel})^2:=-(V_0+2i\epsilon_n^\text{pert,Borel}(i)).
\end{aligned}
\end{equation}
We expect that $\omega_n^\text{Borel}$ gives the \textit{exact} value of the quasinormal frequency. 

We have the perturbative data up to the 200th order.
To perform the Borel summation, we need the analytically continued Borel transform $\cB^\cC[\epsilon_n^\text{pert}](\zeta)$.
How do we get it from these finite data?
Probably, the Pad\'e approximant is the best solution.%
\footnote{Note again that in~\cite{matyjasek2017}, the Pad\'e approximant was used for the original WKB series, which is divergent. 
Here we use the Pad\'e approximant of the Borel transform, which is convergent. As far as we know, the convergency of Pad\'e approximants is guaranteed for convergent series, but we are not sure whether it is so for divergent series.}
Let $f^{[M/N]}(z)$ be the Pad\'e approximant, with an order-$M$ numerator and an order-$N$ denominator, of a given function $f(z)$. %To fix $f^{[M/N]}(z)$ uniquely, one needs the Taylor expansion of $f(z)$ up to $z^{M+N}$.
It is well-known that the Pad\'e approximant works even outside the convergence circle, and moreover captures the singularity structure of the original function.%
\footnote{Since Pad\'e approximants are rational functions, they never have branch point singularities. Nevertheless, Pad\'e approximants tell us about branch cuts. A branch cut appears as a cluster of poles. Consider the Pad\'e approximant of $\log(1+z^2)$, for instance.}
Because of this nice property, the Pad\'e approximant is suitable for our purpose.
Finally, we perform the Laplace transform \eqref{eq:E-Borel-sum} by replacing $\cB^\cC[\epsilon_n^\text{pert}](\hbar \zeta)$ with its Pad\'e $\cB^{[M/N]}[\epsilon_n^\text{pert}](\hbar \zeta)$:
\begin{equation}
\begin{aligned}
\epsilon_n^{\text{pert}, [M/N]}(\hbar)&:=\int_0^\infty d\zeta\, e^{-\zeta} \cB^{[M/N]}[\epsilon_n^\text{pert}](\hbar \zeta), \\
(\omega_n^{[M/N]})^2&:=-(V_0+2i\epsilon_n^{\text{pert}, [M/N]}(i)).
\end{aligned}
\end{equation}
This practical prescription is sometimes called \textit{Borel--Pad\'e summation}.

\begin{figure}[tb]
\begin{center}
  \begin{minipage}[b]{0.4\linewidth}
    \centering
    \includegraphics[width=0.9\linewidth]{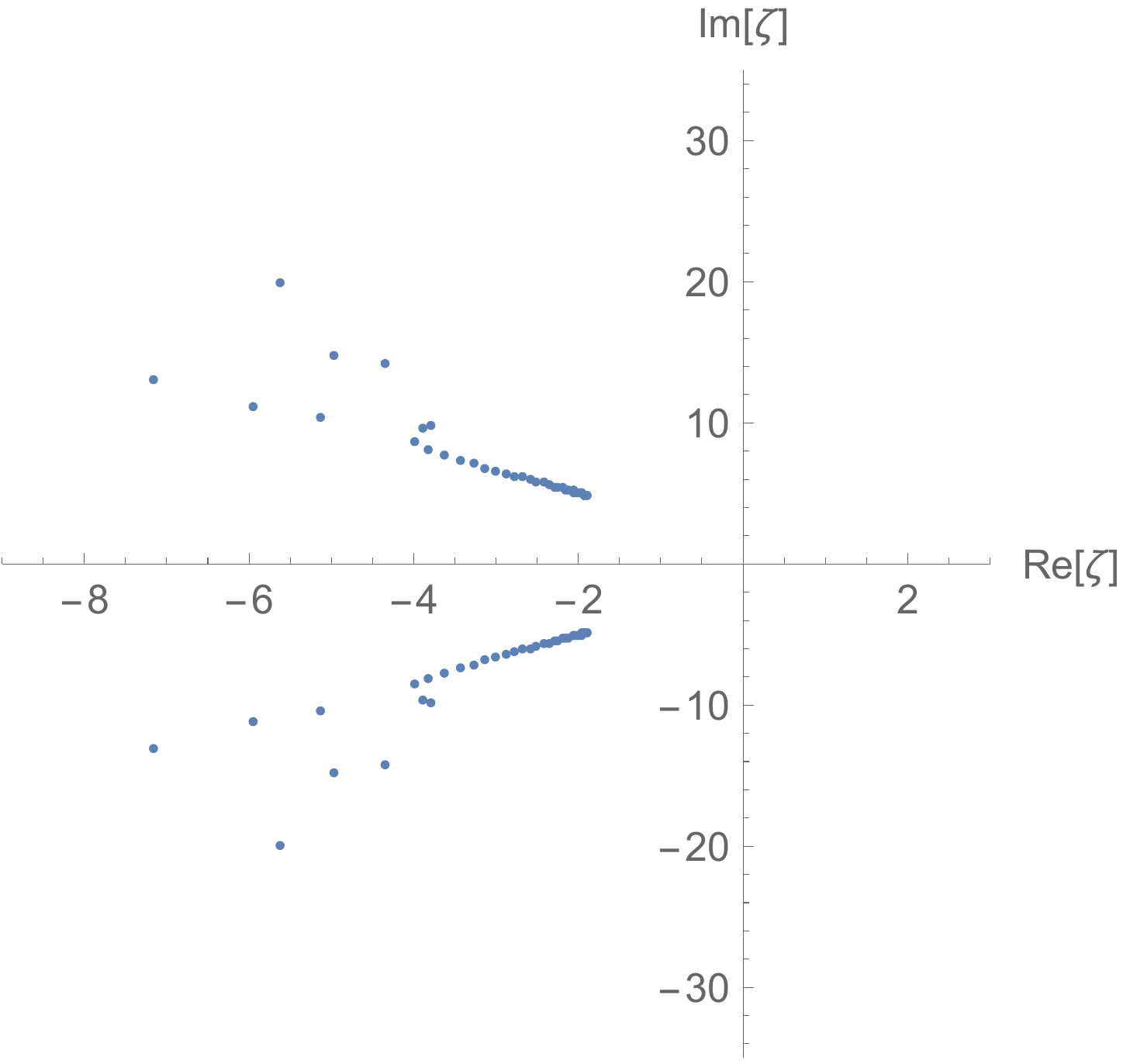}
  \end{minipage} \hspace{0.5truecm}
  \begin{minipage}[b]{0.4\linewidth}
    \centering
    \includegraphics[width=0.9\linewidth]{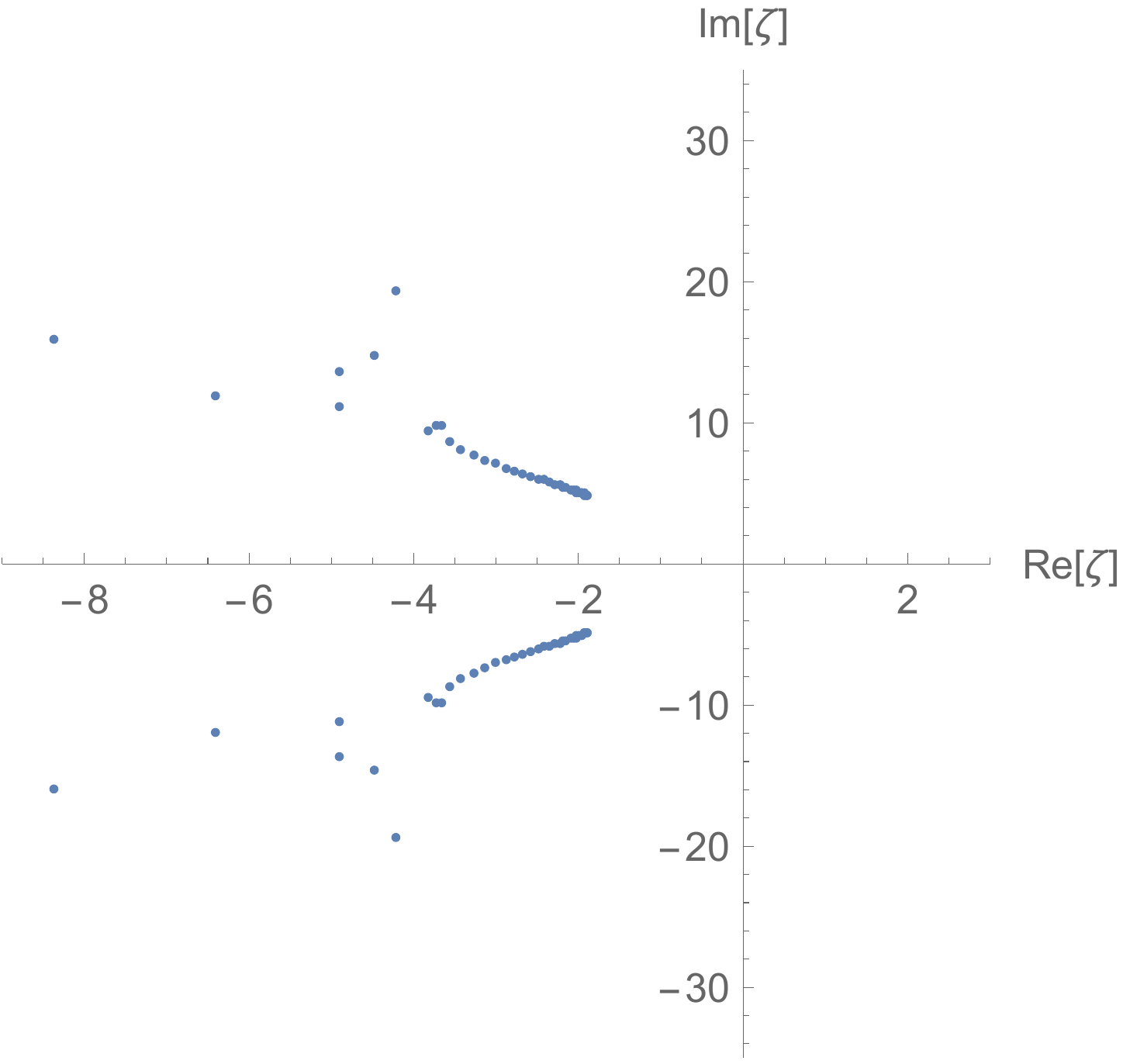}
  \end{minipage} 
\end{center}
  \caption{The pole distributions of the Pad\'e approximants $\cB^{[100/100]}[\epsilon_n^\text{pert}](\zeta)$ for $n=0$ (left) and for $n=1$ (right). There are no singularities on the positive imaginary axis in both cases. A cluster of poles implies a branch cut.}
  \label{fig:sing}
\end{figure}

In the computation of the quasinormal frequencies, we have to do the analytic continuation $\hbar=i$.
The integrand of the Laplace transform becomes $e^{-\zeta} \cB^{[M/N]}[\epsilon_n^\text{pert}](i \zeta)$.
Therefore it is important to see the singularities of $\cB^{[M/N]}[\epsilon_n^\text{pert}](\zeta)$ on the positive \textit{imaginary} axis.
In figure~\ref{fig:sing}, we show the pole structure of the Pad\'e approximant $\cB^{[100/100]}[\epsilon_n^\text{pert}](\zeta)$ with $n=0,1$.
We do not find any singularities on the real axis or on the imaginary axis.
We conclude that the perturbative expansion \eqref{eq:E-pert} is Borel summable both for $\hbar \in \mathbb{R}$ and for $\hbar \in i\mathbb{R}$. 
We play the same game for other values of $l$.
The Borel--Pad\'e summed quasinormal frequencies are shown in table \ref{tab:Borel-QNM}.
These values are compared with known results obtained by other methods.
We have checked that ours are in excellent agreement with the values obtained by Leaver's method~\cite{leaver1985}.

\begin{table}[tb]
\caption{The Borel--Pad\'e summed quasinormal frequencies in the odd-parity gravitational perturbations of the Schwarzschild black hole. We have computed the perturbative expansion of $\epsilon_n$ up to the 200th order, and have used the (diagonal) Pad\'e approximant of the Borel transform. We show only the reliable stable parts of the numerical values. These values are consistent with all the available data up to now in the literature.}
\begin{center}
\begin{tabular}{cccc}
\hline
$l$ & $n$ & $\omega_n^{[100/100]}$ & $|\omega_n^{[100/100]}-\omega_n^{[99/99]}|$ \\
\hline
$2$ & $0$ & $0.74734336883608367159 - 0.17792463137787139656i$ & $ 6.4\times 10^{-24}$\\
      & $1$ & $0.693421993758327 - 0.547829750582470i$ & $1.3 \times 10^{-16}$ \\
      & $2$ & $0.602106909 - 0.956553967 i$ & $8.0 \times 10^{-11} $\\
\hline
$3$ & $0$ & $1.19888657687498014548 - 0.18540609588989520794i$ & $4.8 \times 10^{-42}$\\
      & $1$ & $1.16528760606659886123 - 0.56259622687008808936i$ & $5.0 \times 10^{-34}$ \\
      & $2$ & $1.10336980155690263277 - 0.95818550193392446993i$ & $8.6 \times 10^{-26} $\\
\hline
$4$ & $0$ & $1.61835675506447828139 - 0.18832792197784649881i$ & $7.0 \times 10^{-53}$\\
      & $1$ & $1.59326306406900503032 - 0.56866869880968143729i$ & $1.2 \times 10^{-44}$ \\
      & $2$ & $1.54541906521341859968 - 0.95981635024232615560i$ & $8.0 \times 10^{-37} $\\
\hline
\end{tabular}
\end{center}
\label{tab:Borel-QNM}
\end{table}%

It is easy to see in table~\ref{tab:Borel-QNM} that the convergence speed of the Borel--Pad\'e summation gets better for larger multipole numbers $l$ and smaller overtone numbers $n$.
This is because our method zooms in the bottom of the inverted potential, and it in general works well for a potential with a deep well. 
This is a general property in perturbation theory.

If one considers the scalar perturbations ($s=0$), things get worse.
In the case of $l=0$, we get the following slowly convergent value even for $n=0$:
\begin{equation}
\begin{aligned}
\omega_0^{[100/100]}(s=0, l=0) &\approx 0.220910 - 0.209791i, \\
|\omega_0^{[100/100]}-\omega_0^{[99/99]}| &\approx  1.9 \times 10^{-7}.
\end{aligned}
\end{equation}
The precision should be improved by computing further higher order corrections.

\subsection{Reissner--Nordstr\"om black hole}
The application to the Reissner-Nordstr\"om black hole is almost trivial.
In this case, the (odd-parity) potential is given by
\begin{equation}
\begin{aligned}
V(r_*)=\(1-\frac{1}{r}+\frac{Q^2}{r^2}\)\(\frac{l(l+1)}{r^2}-\frac{q_{2s-3}}{r^3}+\frac{4Q^2}{r^4} \) \qquad (s=1,2),
\end{aligned}
\end{equation}
where $s=1$ is the electromagnetic perturbation, and $s=2$ is the gravitational perturbation.
As in the Schwarzschild case, the mass is set to be $2M=1$.
The parameters $q_{\pm 1}$ are defined by
\begin{equation}
\begin{aligned}
q_{\pm 1}=\frac{1}{2}\Bigl(3 \pm \sqrt{9+16Q^2(l-1)(l+2)}\Bigr).
\end{aligned}
\end{equation}
The tortoise variable is given by
\begin{equation}
\begin{aligned}
r_*=r+\frac{1}{r_+-r_-}\Bigl(r_+^2\log(r-r_+)-r_-^2 \log(r-r_-) \Bigl),
\end{aligned}
\end{equation}
where
\begin{equation}
\begin{aligned}
r_{\pm}:=\frac{1\pm \sqrt{1-4Q^2}}{2}.
\end{aligned}
\end{equation}
In the limit $Q \to 0$, the potential reduces to the Schwarzschild case.
In the limit $Q \to 1/2$, the black hole becomes extremal.
Since $r_+=r_-=1/2$ in this limit, we have to use the following expression:
\begin{equation}
\begin{aligned}
r_*=r-\frac{1}{2(2r-1)}+\log\(r-\frac{1}{2} \).
\end{aligned}
\end{equation}
The remaining computation is completely the same as the non-extremal case.
Note that Leaver's method~\cite{leaver1990} does not work in the extremal case. 
One needs a modification ~\cite{onozawa1996}.
Our approach has no difficulties in this limit.

\begin{table}[tb]
\caption{The Borel--Pad\'e summed quasinormal frequencies with $l=2$ in the odd-parity gravitational perturbations of the Reissner--Nordstr\"om black hole. 
We show only the reliable stable parts of the numerical values. These values are consistent with all the available data up to now in the literature. The case of $Q=1/2$ corresponds to the extremal black hole.}
\begin{center}
\begin{tabular}{cccc}
\hline
$Q$ & $n$ & $\omega_n^{[100/100]}$ & $|\omega_n^{[100/100]}-\omega_n^{[99/99]}|$ \\
\hline
$1/5$ & $0$ & $0.75687377575512748956-0.17879622799342458148 i$ & $1.3 \times 10^{-23}$\\
         & $1$ & $0.7034550613895872 - 0.5502486075914265i$ & $3.2 \times 10^{-17}$ \\
         & $2$ & $0.6128486169 - 0.9598808996 i$ & $6.3 \times 10^{-11} $\\
\hline
$2/5$ & $0$ & $0.80243437092980394128 - 0.17928645946651097029i$ & $4.1 \times 10^{-26}$\\
         & $1$ & $0.7538080116413598975 - 0.5498867488319450640i$ & $4.4 \times 10^{-20}$ \\
         & $2$ & $0.67034301306 - 0.95312777448i$ & $1.2 \times 10^{-12} $\\
\hline
$1/2$ & $0$ & $0.86268160133626430526 - 0.16692063025442518232i$ & $2.0 \times 10^{-30}$\\
      & $1$ & $0.80904710737396060575 - 0.50996873917220399060i$ & $8.2 \times 10^{-26}$ \\
      & $2$ & $0.706802545373 - 0.882745206954i$ & $1.1 \times 10^{-13} $\\
\hline
\end{tabular}
\end{center}
\label{tab:Borel-QNM-RN}
\end{table}%

We repeat all the computations that are done in the Schwarzschild case.
The Borel transform does not have any singularities on the positive imaginary axis again.
The Borel--Pad\'e sums for the $l=2$ mode in the gravitational perturbations with $Q=1/5, 2/5, 1/2$ are shown in table~\ref{tab:Borel-QNM-RN}. 
We have compared these numerical results with the known available data in~\cite{leaver1990, andersson1993, onozawa1996, matyjasek2017}, 
and find perfect agreement.

\section{Summary}\label{sec:summary}
In this paper, we propose a simple procedure to get the quasinormal frequencies of spherically symmetric black holes by combining two old ideas in~\cite{blome1984, ferrari1984, Ferrari:1984zz, bender1969}.
Our recipe consists of the following four steps:
\begin{enumerate}
\setlength{\parskip}{0truemm}
\setlength{\itemsep}{0.8truemm}
\item Compute the Taylor expansion of a given potential in the master equation \eqref{eq:Sch-eq}.
\item Compute the perturbative series of the bound state energy in the inverted potential.
\item Sum it up by the Borel(--Pad\'e) summation method.
\item Do the analytic continuation $\hbar \to i$ at the end.
\end{enumerate}
The second step is most non-trivial, but the recent Mathematica package in~\cite{sulejmanpasic2018} automatically does it!
The high-order computation revealed that the perturbative series of the quasinormal frequencies are divergent and more importantly Borel summable. This strongly supports that the Borel summed frequency $\omega_n^\text{Borel}$ gives the exact quasinormal frequency. We indeed confirmed that the Borel--Pad\'e summation precisely reproduces all the known results. 
Though we showed the explicit computations only for the Schwarzschild and for the Reissner--Nordstr\"om black holes,
our procedure must be widely applicable, as the WKB approach is so.
For instance, we can apply our method to deformations of the Schwarzschild potential~\cite{cardoso2019}.
In fact, we have already confirmed that Table I in \cite{cardoso2019} is easily reproduced very accurately in our framework.

Recently, it was pointed out in \cite{batic2018, batic2019} that there are exact quasi-normal modes that are \textit{not} captured by Leaver's method. Our method in this paper, however, does not reproduce these new modes. Our result rather supports the claim in \cite{panossomacedo2019}.
If the claim in \cite{batic2018, batic2019} is correct, their new quasi-normal modes should not have any counterparts of bound states.
It would be interesting to understand their new modes on the inverted potential side.

In this work, we focused on the behavior near the top (or the bottom) of the potential (or the inverted potential).
This restriction makes it hard to obtain the accurate values of the quasinormal frequencies $\omega_n$ with high overtone numbers $n$.
In addition, the convergence of the Borel--Pad\'e summation of $\omega_n$ in the scalar perturbations is quite slow because of the shallow well of the inverted potential.
These difficulties are probably resolved in the approach in~\cite{froman1992}.
In bound state problems, the energy spectrum is approximately obtained by Bohr--Sommerfeld quantization conditions. These quantization conditions are improved by taking into account high-order quantum corrections~\cite{dunham1932}.%
\footnote{However, sometimes they receive nonperturbative corrections \cite{balian1978, voros1983}.}
The same is true for quasinormal modes. Quantum corrected Bohr--Sommerfeld conditions are found in~\cite{froman1992}.
However, even for the Schwarzschild black hole, their exact forms have not been known.
It would be significant to discuss the Borel analysis for the quantization conditions in~\cite{froman1992}.

Finally, our method heavily relies on the assumption \eqref{eq:QNM-BS}. Though we have tested it for a few examples,
it is important to (dis)prove the assumption \eqref{eq:QNM-BS} rigorously. 
As already mentioned before, we have to care about the Stokes phenomenon.
This problem should be solved by understanding an evolution of (anti-)Stokes curves in the $r_*$-plane (or in the $r$-plane) from $\arg \hbar=0$ to $\arg \hbar=\pi/2$. The similar evolution in the pure quartic oscillator is found in \cite{voros1983}.

\acknowledgments{This work was inspired by a nice seminar on ``Parametrized Black Hole Quasinormal Ringdown'' by Masashi Kimura at Rikkyo University. His talk was clear for non-experts, and drew my attention to this fascinating topic. I also thank him for telling me a lot of basics and references on the quasinormal modes. This work is supported by JSPS KAKENHI Grant Number JP18K03657.}

\appendix

\section{Review of Borel summation method}\label{app:Borel}
In this appendix, we briefly review the Borel summation method.
This method is important in physics because we can probe nonperturbative corrections from perturbative series.
We refer the reader to an educational lecture note~\cite{marino2014} and a comprehensive review~\cite{aniceto2019} on this topic.

Let $f_\text{pert}(g)$ be a formal perturbative series of a given function $f(g)$:
\begin{equation}
\begin{aligned}
f(g) \simeq f_\text{pert}(g)=\sum_{k=0}^\infty f_k g^k,
\end{aligned}
\label{eq:f-pert}
\end{equation}
where $\simeq$ means that both sides are equal in the asymptotic sense.
This point is not crucial in our analysis.
We assume that the sequence $f_k$ satisfies the following condition:
\begin{equation}
\begin{aligned}
|f_k| \leq M C^k k!,
\end{aligned}
\label{eq:Gevrey-1}
\end{equation}
where $M$ and $C$ are constants.
Mathematically, the series \eqref{eq:f-pert} with the condition \eqref{eq:Gevrey-1} are called Gevrey-1 series.
Most perturbative series in physics belong to this class.
Therefore throughout this appendix, we consider only Gevrey-1 series.
Gevrey-1 series are in general divergent series.
For divergent series, there exist several functions that have the same perturbative series because of the Stokes phenomenon.
It is far from obvious to reconstruct the exact function $f(g)$ from its divergent perturbative series $f_\text{pert}(g)$.
The Borel summation provides us a hint on this problem.

We first define the Borel transform of \eqref{eq:f-pert} by
\begin{equation}
\begin{aligned}
\cB[f_\text{pert}](\zeta):=\sum_{k=0}^\infty \frac{f_k}{k!} \zeta^k.
\end{aligned}
\end{equation}
Due to the Gevrey-1 condition \eqref{eq:Gevrey-1}, the Borel transform has a \textit{finite} radius of convergence.
The radius of convergence is determined by the nearest singularity of the Borel transform from the origin in the complex $\zeta$-plane (sometimes called the Borel plane).
Note that if $f_\text{pert}(g)$ is a convergent series, then its Borel transform is an entire function and has no singularities (except for $\zeta=\infty$). Inverting the logic, the existence of singularities in the Borel plane means that the sequence 
$f_k$ factorially diverges in $k\to \infty$.

The Borel transform $\cB[f_\text{pert}](\zeta)$ is analytically continued outside the convergence circle.
We denote it by $\cB^\cC[f_\text{pert}](\zeta)$.
The Borel sum is finally defined by the Laplace transform
\begin{equation}
\begin{aligned}
f_\text{pert}^\text{Borel}(g):=\int_0^\infty d\zeta \, e^{-\zeta} \cB^\cC[f_\text{pert}](g \zeta).
\end{aligned}
\label{eq:Borel-sum}
\end{equation}
The Borel sum \eqref{eq:Borel-sum} has the same asymptotic perturbative expansion of the original series \eqref{eq:f-pert}
because the factorial $k!$ has the following integral representation:
\begin{equation}
\begin{aligned}
k!=\int_0^\infty d\zeta \, e^{-\zeta} \zeta^k.
\end{aligned}
\end{equation}
Does the Borel sum $f_\text{pert}^\text{Borel}(g)$ agree with the exact function $f(g)$?
Sometimes, the answer is yes, but in general it is not true. 
The reason is as follows.

We suppose $g>0$. The Borel transform $\cB^\cC[f_\text{pert}](\zeta)$ has singularities in the Borel plane.
If the Laplace transform in \eqref{eq:Borel-sum} is well-defined, then the series \eqref{eq:f-pert} is called \textit{Borel summable}.
Sometimes, some singularities of $\cB^\cC[f_\text{pert}](\zeta)$ are located on the positive real axis, and the Laplace transform in \eqref{eq:Borel-sum} is not defined.
This case is called \textit{Borel non-summable}. 
In the Borel non-summable case, one has to deform the integration contour to avoid these singularities. There are two possible directions to do so, and one can define two modified Borel sums $f_\text{pert}^{\pm \text{Borel}}(g)$ for these deformed contours.
Importantly, these modified Borel sums $f_\text{pert}^{\pm \text{Borel}}(g)$ have the same perturbative expansion \eqref{eq:f-pert}, but they have an exponentially small difference:
\begin{equation}
\begin{aligned}
f_\text{pert}^{+\text{Borel}}(g)=f_\text{pert}^{-\text{Borel}}(g)+\cO(e^{-A/g}).
\end{aligned}
\label{eq:discon}
\end{equation}
The difference is \textit{nonperturbative} in $g$. 
The magnitude $A$ is determined by the nearest singularity of the Borel transform from the origin~\cite{marino2014}.
The discontinuity \eqref{eq:discon} is nothing but the Stokes phenomenon in the Borel summation.
The positive real axis is the Stokes line.

Quite interestingly, such an ambiguity is removed by taking into account
nonperturbative corrections $f_\text{np}(g)\sim \cO(e^{-A/g})$ to the original perturbative series.
Then the exact function is reconstructed by the combination of the perturbative series and the nonperturbative corrections without the ambiguity.
Very roughly, we have
\begin{equation}
\begin{aligned}
f(g)=(f_\text{pert}+C_+ f_\text{np})^{+\text{Borel}}(g)=(f_\text{pert}+C_- f_\text{np})^{-\text{Borel}}(g),
\end{aligned}
\label{eq:exact-f}
\end{equation}
where $C_\pm$ are Stokes constants. The asymptotic expansion now takes the form like
\begin{equation}
\begin{aligned}
f(g) \simeq \sum_{k=0}^\infty f_k g^k+C_{\pm} e^{-A/g}\sum_{k=0}^\infty f_k^{(1)} g^k+\cdots
\end{aligned}
\label{eq:f-trans}
\end{equation}
where $\cdots$ means the higher order nonperturbative corrections.
The asymptotic expansion \eqref{eq:f-trans} is called the \textit{transseries expansion}.

In summary, if a perturbative expansion is Borel summable, one can expect that its Borel sum agrees with the exact function.%
\footnote{However, there is a counterexample of this statement in string theory \cite{grassi2015}.}
The analysis in the main text corresponds to this case.
If a perturbative expansion is Borel non-summable, there is an ambiguity to define modified Borel sums.
We have to add nonperturbative corrections to cancel out the ambiguity.
The perturbative sector and the nonperturbative sector are interrelated in a non-trivial way.
In fact, the equations \eqref{eq:discon} and \eqref{eq:exact-f} can be regarded as a constraint to nonperturbative corrections if we know the perturbative part.
This is why the Borel summation is important both in mathematics and in physics. See \cite{marino2014, aniceto2019} in more detail.

\bibliographystyle{amsmod}
\bibliography{QNM}

\end{document}